\documentclass[12pt]{article}
\usepackage{epsfig}
\usepackage{amssymb}
\usepackage{amsmath}
\usepackage{amsfonts}
\usepackage{graphicx}
\usepackage{mathrsfs}
\DeclareMathAlphabet{\mathscrbf}{OMS}{mdugm}{b}{n}
\usepackage{mathabx}
\usepackage[dvips]{color}
\usepackage{multirow}
\usepackage{calc}
\usepackage{accents}

\newcommand{\R}{\mathbb{R}}
\newcommand{\C}{\mathbb{C}}

\newcommand{\fz}{\mathfrak{z}}

\newcommand{\bk}{\mathbf{k}}

\newcommand{\bp}{{\mathbf{p}}}
\newcommand{\bfr}{\mathbf{r}}

\newcommand{\bx}{\mathbf{x}}
\newcommand{\by}{\mathbf{y}}
\newcommand{\bu}{\mathbf{u}}

\newcommand{\bE}{\mathbf{E}}

\newcommand{\bF}{\mathbf{F}}

\newcommand{\bI}{\mathbf{I}}

\newcommand{\cR}{\mathcal{R}}

\newcommand{\cZ}{\mathcal{Z}}
\newcommand{\be}{\begin{equation}}
\newcommand{\ee}{\end{equation}}
\newcommand{\bea}{\begin{eqnarray}}
\newcommand{\eea}{\end{eqnarray}}
\newcommand{\nn}{\nonumber}

\newcommand{\ed}{\end{document}}

\newcommand{\np}{\newpage}

\newcommand{\bi}{\begin{itemize}}
\newcommand{\ei}{\end{itemize}}

\newcommand{\bce}{\begin{center}}
\newcommand{\ece}{\end{center}}

\newcommand{\sD}{\mathscr{D}}

\newcommand{\sM}{\mathscr{M}}

\newcommand{\sR}{\mathscr{R}}

\newcommand{\RE}{{\rm Re}}
\newcommand{\IM}{{\rm Im}}

\newcommand{\bzero}{{\boldsymbol{0}}}

\newcommand{\for}{{\mbox{\rm for}}}

\newcommand{\bfeta}{{\boldsymbol{\eta}}}

\newcommand{\bfvarepsilon}{{\boldsymbol{\varepsilon}}}
\newcommand{\bfmu}{{\boldsymbol{\mu}}}

\begin{document}

\title{Broadband directional invisibility}


\author{Farhang Loran\thanks{E-mail address: loran@iut.ac.ir}
~and Ali~Mostafazadeh\thanks{Corresponding author, E-mail address:
amostafazadeh@ku.edu.tr}\\[6pt]
$^*$Department of Physics, Isfahan University of Technology, \\ Isfahan 84156-83111, Iran\\[6pt]
$^\dagger$Departments of Mathematics and Physics, Ko\c{c} University,\\  34450 Sar{\i}yer,
Istanbul, T\"urkiye}

\date{ }
\maketitle

\begin{abstract}
The discovery of unidirectional invisibility and its broadband realization in optical media satisfying spatial Kramers-Kronig relations are important landmarks of non-Hermitian photonics. We offer a precise characterization of a higher-dimensional generalization of this effect and find sufficient conditions for its realization in the scattering of scalar waves in two and three dimensions and electromagnetic waves in three dimensions.  More specifically, given a positive real number $\alpha$ and a continuum of unit vectors $\Omega$, we provide explicit conditions on the interaction potential (or the permittivity and permeability tensors of the scattering medium in the case of electromagnetic scattering) under which it displays perfect (non-approximate) invisibility whenever the incident wavenumber $k$ does not exceed $\alpha$ (i.e., $k\in(0,\alpha]$) and the direction of the incident wave vector ranges over $\Omega$. A distinctive feature of our approach is that it allows for the construction of potentials and linear dielectric media that display perfect directional invisibility in a finite frequency domain.
\end{abstract}

The study of scattering systems that display invisibility has been an important subject of research for decades. There are different mechanisms that render a scattering system invisible \cite{Fleury-2014}. Among these is the presence of regions in a scattering medium with specific gain and loss profiles \cite{lin-2011,pra-2013}.  A particularly remarkable development in this direction is the discovery, in one dimension, that the potentials $v(x)$ whose Fourier transform $\tilde v(p)$ vanishes on the negative or positive half-line display unidirectional invisibility for all values of the incident wavenumber $k$, \cite{horsley-2015,longhi-2015,horsley-longhi,jiang-2017,ye-2017,zhang-2021,zheng-2023}. In terms of the scattered wave vector $\bk_{\rm s}$, the incident wave vector $\bk_{\rm i}$, and the scattering amplitude $f(\bk_{\rm s},\bk_{\rm i})$ of the potential \cite{tjp-2020}, we can state the unidirectional invisibility of these potentials in the form\footnote{In one dimension, we respectively identify the wave vectors $\bk_{\rm s}$ and $\bk_{\rm s}$ with the real numbers $k_{\rm s}$ and $k_{\rm i}$ satisfying  $\bk_{\rm s}=k_{\rm s}\hat\bx$ and $\bk_{\rm s}=k_{\rm s}\hat\bx$, where $\hat\bx$ is the unit vector joining 0 to 1 on the real line.}: 
	\begin{align}
	&f(\bk_{\rm s},\bk_{\rm i})=0~~{\rm for}~~
	\bk_{\rm i}<0~{\rm or}~\bk_{\rm i}>0.
	\label{condi}
	\end{align}
Because $|\bk_{\rm s}|=|\bk_{\rm i}|=k$,  we can also express (\ref{condi}) in the form
	\begin{align}
	&f(\bk_{\rm s},\bk_{\rm i})=0~~\for~~k\in\R^+~{\rm and}~~\hat\bk_{\rm i}=- 1~{\rm or}~1,
	\label{condi1}
	\end{align}
where $\R^+$ is the set of positive real numbers, and $\hat\bk_{\rm i}:=k^{-1}\bk_{\rm i}$. A natural multi-dimensional extension of (\ref{condi1}) is
	\begin{align}
	&f(\bk_{\rm s},\bk_{\rm i})=0~~{\rm for}~~k\in \cR~~{\rm and}~~\hat\bk_{\rm i}\in \Omega,
	\label{condi2}
	\end{align}
where $\cR$ is a given subset of $\R^+$, $\Omega$ is a given subset of the set $S^{d-1}$ of unit vectors in the Euclidean space $\R^d$, and $d$ is the spatial dimension. We can also consider the following generalizations of (\ref{condi2}). 
	\begin{align}
	&f(\bk_{\rm s},\bk_{\rm i})=0~~{\rm for}~~\bk_{\rm i}\in {Q},
	\label{condi2-gen}\\
	&f(\bk_{\rm s},\bk_{\rm i})=0~~{\rm for}~~\bk_{\rm i}\in {Q}~~{\rm and}~~
	\hat\bk_{\rm s}\in \Omega_{\rm s},
	\label{condi2-gen2}
	\end{align}
where ${Q}$ and $\Omega_{\rm s}$ are respectively given subsets of $\R^d$ and $S^{d-1}$. Relation~(\ref{condi2-gen2})  with $\bk_{\rm i}\in {Q}$ changed to $\bk_{\rm s}-\bk_{\rm i}\in {Q}$ corresponds to the ``invisibility on demand'' considered in Ref.~\cite{hayran-2018a}. Strictly speaking potentials fulfilling (\ref{condi2-gen2}) will be invisible if and only if (\ref{condi2-gen}) holds for all $\hat\bk_{\rm s}\in S^{d-1}$, i.e., (\ref{condi2-gen2}) holds for $\Omega_{\rm s}= S^{d-1}$. 

Condition (\ref{condi2}) corresponds to the invisibility of the scattering medium for incident waves with wavenumber in the range $\cR$ and incidence direction belonging to $\Omega$. If $\cR$ is an interval in $\R^+$, this  is a broadband invisibility. If $\Omega\neq S^{d-1}$, it is a direction-dependent or directional invisibility. The main purpose of the present article is to identify a class of scattering systems that display broadband directional invisibility.

In one dimension, the vanishing of the Fourier transform of a potential $v(x)$ on the negative or positive half-line implies that its real and imaginary parts are related by the Hilbert transformation. This in turn shows that they satisfy spatial Kramers-Kronig relations \cite{horsley-2015}. The authors of Refs.~\cite{hayran-2018a,hayran-2018b} propose achieving ``invisibility on demand'' by considering potentials whose real and imaginary parts are linked by a generalization of the Hilbert transform \cite{Ahmed-2018,Ahmed-2020}. This method relies on an argument borrowed from Ref.~\cite{horsley-2015} which reduces the condition of the reflectionlessness of a scattering medium to the vanishing of the scattered field calculated using the first Born approximation. For future reference we reproduce this argument in the sequel.

First, we recall that the Fourier transform\footnote{We use convenstions where the Fourier transform of a function $\phi(x)$ is given by $\tilde \phi(p):=\int_{-\infty}^\infty dx\: e^{-ipx}\phi(x)$.} of the terms {$e_{\rm s}^{(n)}$} in the Born series for the scattered field,  $e_{\rm s}(x)=\sum_{n=1}^\infty e_{\rm s}^{(n)}(x)$, are given by
	\begin{align}
	\tilde e_{\rm s}^{(1)}(p)&=\frac{1}{{2\pi}}G(p)\tilde v(p-\bk_{\rm i}),
	\label{eq1}\\
	\tilde e_{\rm s}^{(n+1)}(p)&=\frac{1}{{2\pi}}G(p)\int_{-\infty}^\infty
	dq\:\tilde v(p-q)\tilde e_{\rm s}^{(n)}(q),
	\label{eq2}
	\end{align}
where $G(p):=(k^2-p^2+i\epsilon)^{-1}$.
Suppose that $\bk_i>0$ and 
	\be
	\tilde v(p)=0~~\for~~p<0.
	\label{Hardy}
	\ee 
Then (\ref{eq1}) implies that $\tilde e_{\rm s}^{(1)}(p)=0$ for $p<0$, because in this case $p-\bk_{\rm i}<0$ and $\tilde v(p-\bk_{\rm i})=0$. For $n=1$, this shows that the integral on the right-hand side of (\ref{eq2}) is to be evaluated over $\R^+$, i.e., for $q>0$. But then for $p<0$, the integrand vanishes and we find $\tilde e_{\rm s}^{(2)}(p)=0$ for $p<0$. This argument applies for $n\geq 2$ and gives $\tilde e_{\rm s}^{(n>2)}(p)=0$ for $p<0$. Therefore, $\tilde e_{\rm s}(p)=0$ for $p<0$, i.e., the scattered wave has no left-going Fourier modes. This means that there is no reflected wave. For a short-range potential \cite{yafaev}, $\int_{-\infty}^\infty dx\,|v(x)|<\infty$, and it turns out that the incident waves do not get affected upon transmission either \cite{longhi-2015}. Hence they satisfy the (uni)directional invisibility condition (\ref{condi1}) with $\hat\bk_{\rm i}=+1$. 

The first part of the above argument establishes $\tilde e_{\rm s}^{(1)}(p)=0$ for $p<0$ and the second part shows that this together with (\ref{Hardy}) imply $\tilde e_{\rm s}^{(n>1)}(p)=0$ for all $p<0$. Motivated by this argument and the two-dimensional generalization of (\ref{eq1}), the authors of Ref.~\cite{hayran-2018a} maintain that in order to achieve ``invisibility on demand'' in two dimensions (2D) it is sufficient to ensure that the (two-dimensional) Fourier transform of the potential vanishes on a set of wave vectors. This claim relies on the assertion that ``when $e_{\rm s}^{(1)}$ (first-order Born approximation) is completely suppressed, then every successive order will also be zero'' \cite{hayran-2018a}. This is clear from (\ref{eq2}) provided that one interprets ``when $e_{\rm s}^{(1)}$ $\cdots$ is completely suppressed'' as ``$e_{\rm s}^{(1)}(p)=0$ for all $p\in\R$.'' The difficulty with this requirement is that, in view of (\ref{eq1}), it is equivalent to $\tilde v(p)=0$ for all $p\in\R$ which holds if and only if $v(x)=0$ for all $x\in\R$. If $\tilde e_{\rm s}^{(1)}(p)\neq 0$ in a (non-discrete) subset of $\R$, the higher order terms in the Born series need not vanish. The approach of Refs.~\cite{hayran-2018a,hayran-2018b} is therefore applicable only for cases where the first Born approximation is valid. It is well-known that in this case, inverse Fourier transformation of the desired scattering data offers a simple but approximate solution for the general inverse scattering problem \cite{Wolf,Devaney,Carter,Chadan-IS}. This method can in particular be applied to construct potentials that display approximate directional invisibility. A concrete example is the class of unidirectionally invisible potentials of Ref.~\cite{prsa-2016}.

The inverse scattering based on the first Born approximation is considered undesirable, because it can be safely employed if the strength of the potential one looks for is much smaller than the energy scale of interest, i.e., $|v(\bfr)|\ll k^2$.  In particular, the results of Refs.~\cite{hayran-2018a,hayran-2018b} hold for incident waves with sufficiently large wavenumbers, i.e., at high frequencies. There are however potentials of arbitrary strength for which the first Born approximation is exact for sufficiently low frequencies \cite{pra-2019,pra-2021}. The development of the dynamical formulation of stationary scattering of scalar waves \cite{pra-2016a,pra-2021} has paved the way for the construction of such potentials in 2D and 3D. In Ref.~\cite{pra-2019} we use them to realize broadband unidirectional invisibility for scalar waves  in 2D, i.e., find potentials that satisfy (\ref{condi2}) for $\cR=(0,\alpha]$, where $\alpha\in\R^+$, and $\Omega$ is a semi-circle. Recently, we have devised a dynamical formulation of electromagnetic scattering \cite{pra-2023} and extended our results on the exactness of the first Born approximation to electromagnetic (EM) waves \cite{p176}. In the remainder of this article, we use the results of Refs.~\cite{pra-2021,pra-2019,p176} on the exactness of the first Born approximation to identify scattering systems displaying broadband directional invisibility for scalar and electromagnetic waves.


We begin our investigation by considering the scattering of scalar waves by a short-range potential $v(x,y)$ in 2D. Let $\alpha\in\R^+$ and suppose that the Fourier transform of $v(x,y)$ with respect to $y$, which we denote by $\tilde v(x,p_y)$, satisfies 
	\be
	\tilde v(x,p_y)=0~~\for~~p_y\leq\alpha.
	\label{condi-2Da}
	\ee
Then the first Born approximation turns out to give the exact expression for the scattering amplitude of this potential for incident waves with wavenumber $k\leq\alpha$, \cite{pra-2021,pra-2019}. Because of the freedom in the choice of the coordinate system, (\ref{condi-2Da}) is equivalent to the existence of a unit vector $\hat\bu$ such that the two-dimensional Fourier transform of the potential, $\tilde v(\bp)$, satisfies
	\be
	\tilde v(\bp)=0~~\for~~\hat\bu\cdot\bp\leq\alpha,
	\label{condi-2Db}
	\ee
where $\bp:=(p_x,p_y)\in\R^2$ is arbitrary. 

Suppose that (\ref{condi-2Db}) holds. Then the first Born approximation is exact for $k\leq\alpha$, and the scattering amplitude is proportional to $\tilde v(\bk_{s}-\bk_{\rm i})$ \cite{pra-2021,pra-2019}. This shows that whenever $\cR\subseteq (0,\alpha]$ we can express the invisibility condition (\ref{condi2}) in the form
	\begin{align}
	&\tilde v(\bk_{\rm s}-\bk_{\rm i})=0~~{\rm for}~~k\in \cR~~{\rm and} ~~ \hat\bk_{\rm i}\in \Omega.
	\label{condi2-tv}
	\end{align}
For incident waves with $k\leq \frac{\alpha}{2}$, we can use (\ref{condi-2Db}) and the inequalities,
	$\hat\bu\cdot(\bk_{\rm s}-\bk_{\rm i})\leq|\bk_{\rm s}-\bk_{\rm i}|\leq 2k\leq \alpha$, to infer that $\tilde v(\bk_{\rm s}-\bk_{\rm i})=0$ for all $\hat\bk_{\rm i}\in S^1$, i.e., $v$ is omnidirectionally invisible \cite{ol-2017}. Consequently, directional invisibility requires that $\cR\subseteq(\frac{\alpha}{2},\mbox{\small$\alpha$}]$.\footnote{The broadband unidirectional invisibility discussed in Ref.~\cite{pra-2019} is a particular example with $\cR=(\frac{\alpha}{\sqrt 2},\mbox{\small$\alpha$}]$.}
	
Next, we introduce
	\begin{align}
	&\Pi:=\left\{\left.\bp\in\R^2\,\right|\,\hat\bu\cdot\bp\leq\alpha\,\right\},
	\nn\\
	&\Delta:=\left\{\bk_{\rm s}-\bk_{\rm i}\,\left|\,k\in\cR,\,
	\hat\bk_{\rm s}\in S^1,\, \hat\bk_{\rm i}\in \Omega\right.\right\},
	\nn\\
	&\cZ:=\left\{\left.\bp\in\R^2\,\right|\,\tilde v(\bp)=0\,\right\},
	\nn
	\end{align}
and express (\ref{condi-2Db}) and (\ref{condi2-tv}) as  
$\Pi \subseteq \cZ$ and $\Delta \subseteq \cZ$, respectively.\footnote{$\Pi$ is the half plane containing the origin that is bounded by the line $\hat\bu\cdot\bp=\alpha$.} The latter follows from the former, if $\Delta\subseteq\Pi$. Combining this observation with the fact that  (\ref{condi-2Db}) and (\ref{condi2-tv}) imply (\ref{condi2}) for $\cR\subseteq(\frac{\alpha}{2},\alpha]$, we arrive at the following sufficient condition for the invisibility of the potential for $k\in{\cR}\subseteq(\frac{\alpha}{2},\alpha]$ and $\hat\bk_{\rm i}\in\Omega$.
	\be
	\Delta\subseteq\Pi\subseteq\cZ.
	\label{condi-main}
	\ee

Let us try to find potentials that are invisible for wavenumbers ranging over an interval, i.e., $\cR=[\rho_-,\rho_+]$ where $\rho_\pm\in\R^+$ and  $\rho_-<\rho_+$. To do this, we set $\hat\bu:=\hat\by$, where $\hat\by$ is the unit vector along the $y$ axis, and demand that $v$ satisfies (\ref{condi-2Da}) for some $\alpha\geq\rho_+$, so that $\Pi\subseteq\cZ$. For $\rho_+\leq\frac{\alpha}{2}$, $v$ is omnidirectionally invisible for all $k\in\sR$, \cite{ol-2017}. We therefore confine our attention to cases where $\rho_+>\frac{\alpha}{2}$.

It is easy to see that the potentials fulfilling (\ref{condi-2Da}) have the form \cite{ol-2017},
	\be
	v(x,y)=e^{i\alpha\,y}\int_0^\infty dq\, e^{iqy}w(x,q),
	\label{v=1}
	\ee
where $w(x,q)$ is a uniformly continuous function of $q$ and an arbitrary function of $x$ such that the right-hand side of this equation gives a short-range potential.\footnote{This is the case if $\int_0^\infty dq\, e^{iqy}w(x,q)$ exists and tends to zero for $r\to\infty$ faster than $1/r$, where $r:=\sqrt{x^2+y^2}$.} A concrete example is
	\be
	v(x,y)=\frac{\fz\,\chi_a(x)e^{i\alpha\,y}}{\left(y+i\ell\right)^{n+1}},
	\label{eg-2D}
	\ee
where $\fz$ is a real or complex coupling constant, $a$ and $\ell$ are positive real parameters, $n$ is a positive integer, and 
	\[\chi_a(x):=\left\{\begin{array}{ccc}
	1 &\for &x\in[-\frac{a}{2},\frac{a}{2}],\\
	0 &\for &x\notin[-\frac{a}{2},\frac{a}{2}].\end{array}\right.\]
Eq.~(\ref{eg-2D}) follows from Eq.~(\ref{v=1}), if we set $w(x,q)=\varsigma\:\chi_a(x)q^n e^{-\ell q}$ where $\varsigma:=2\pi(-i)^{n+1}\fz/n!$.\footnote{We can obtain more general examples of potentials satisfying (\ref{condi-2Da}) by adding arbitrary number of potentials of the form (\ref{eg-2D}) with different values for $\fz,a,\ell$, and $n$.}

To determine the directions along which the potentials of the form (\ref{v=1}) are guaranteed to display broadband directional invisibility for $k\in\cR:=[\rho_-,\rho_+]$, we note that because $\Pi\subseteq\cZ$, the invisibility condition (\ref{condi-main}) reduces to $\Delta\subseteq\Pi$. This means
	$\hat\bu\cdot(\bk_{\rm s}-\bk_{\rm i})\leq\alpha~~{\rm for}~~k\in \cR~~{\rm and}~~\hat\bk_{\rm i}\in \Omega$.
We therefore wish to find $\Omega$ such that this relation holds. 	

\begin{figure}
        \begin{center}
        \includegraphics[scale=.35]{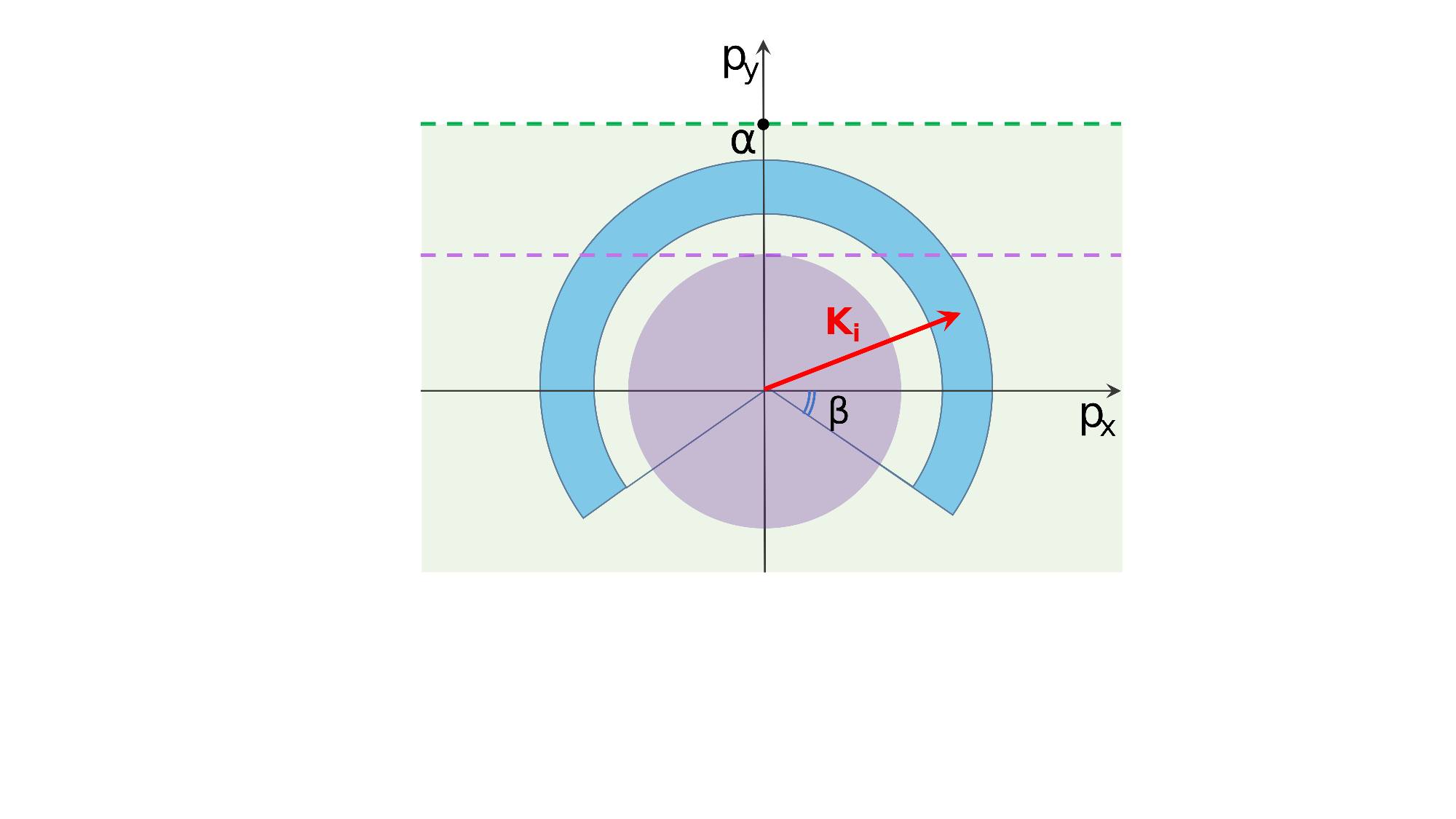}\vspace{-1.5cm}
        \caption{Schematic representation of regions of invisibility of the potentials (\ref{v=1}) for $k\in\cR=[\rho_-,\rho_+]$ and $\frac{\alpha}{2}<\rho_-<\rho_+<\alpha$. The green and purple dashed lines are respectively the graphs of $p_y=\alpha$ and $p_y=\frac{\alpha}{2}$.  The half-plane $\Pi$ and the region of directional invisibility $\Delta$ are painted in light green and blue, respectively. These potentials are omnidirectionally invisible in the purple disc. $\beta:=\arcsin(\frac{\alpha}{\rho_+}-1)$.}
        \label{fig1}
        \end{center}
        \end{figure}

Let $\theta$ and $\theta_0$ be respectively the angles $\bk_{\rm s}$ and $\bk_{\rm i}$ make with the positive $x$ axis. Then the condition ``$\hat\bu\cdot(\bk_{\rm s}-\bk_{\rm i})\leq\alpha$ for all $\hat\bk_{\rm s}\in S^1$'' is equivalent to ``$\sin\theta-\sin\theta_0\leq \frac{\alpha}{k}$ for all $\theta\in[0,360^\circ)$''. The latter holds if and only if $\sin\theta_0\geq 1-\frac{\alpha}{k}$. To satisfy this condition for all  $k\in[\rho_-,\rho_+]$, we must have $\sin\theta_0\geq 1-\frac{\alpha}{\rho_+}$, i.e., $\theta_0\in[-\beta,180^\circ+\beta]$ where $\beta:=\arcsin(\frac{\alpha}{\rho_+}-1)$. This argument identifies $\Omega$ with 
	$\Omega_\beta:=\{(\cos\varphi,\sin\varphi)|-\beta\leq\varphi\leq 180^\circ+\beta\}$.
If $\rho_->\frac{\alpha}{2}$, the potentials (\ref{v=1})  are directionally invisible for all $k\in[\rho_-,\rho_+]$ and $\hat\bk_{\rm i}\in\Omega_\beta$. If $\rho_-\leq\frac{\alpha}{2}$, they are omnidirectionally invisible for $k\in[\rho_-,\frac{\alpha}{2}]$ and directionally invisible for $k\in(\frac{\alpha}{2},\rho_+]$ and $\hat\bk_{\rm i}\in\Omega_\beta$. Figure~\ref{fig1} shows the regions of broadband invisibility of these potentials for $\frac{\alpha}{2}<\rho_-<\rho_+<\alpha$.
        
The above analysis of broadband invisibility has a straightforward extension to 3D. For a short-range potential $v:\R^3\to\C$, the first Born approximation is exact if there is a unit vector $\bu$ in the $x$-$y$ plane such that
	\be
	\tilde v(p_x,p_y,z)=0~~\for~~\hat\bu\cdot\vec p\leq\alpha,
	\label{exact-3D-1}
	\ee
where $\tilde v(p_x,p_y,z)$ stands for the two-dimensional Fourier transform of $v(x,y,z)$ with respect to $x$ and $y$, and $\vec p:=p_x\hat\bx+p_y\hat\by$,\cite{pra-2021}. Clearly, we can express  (\ref{exact-3D-1}) in the form (\ref{condi-2Db}) provided that we identify $\tilde v(\bp)$ with the 3D Fourier transform of $v(\bfr)$, i.e., $\tilde v(\bp):=\int_{\R^3}d^3\!\bfr\, e^{-i\bp\cdot\bfr}v(\bfr)$. Again, in view of the freedom of choice of our coordinate system, we can identify $\bu$ as an arbitrary unit vector in $\R^3$, and define the 3D analogs of $\Pi$, $\Delta$, and $\cZ$ by 
	\begin{align}
	&\Pi:=\left\{\left.\bp\in\R^3\,\right|\,\hat\bu\cdot\bp\leq\alpha\,\right\},
	\label{Pi-def-3D}\\
	&\Delta:=\left\{\bk_{\rm s}-\bk_{\rm i}\,\left|\,k\in\cR,\,
	\hat\bk_{\rm s}\in S^2,\, \hat\bk_{\rm i}\in \Omega\right.\right\},
	\label{Delta-def-3D}\\
	&\cZ:=\left\{\left.\bp\in\R^3\,\right|\,\tilde v(\bp)=0\,\right\}.
	\end{align}
This allows for a direct application of our characterization of broadband directional invisibility to 3D, because the scattering amplitude of the potential computed using the first Born approximation is proportional to $\tilde v(\bp)$, \cite{Sakurai}.
	

Next, we consider the scattering of plane electromagnetic waves of angular frequency $\omega$ by a stationary linear medium having relative permittivity and relative permeability tensors, $\hat\bfvarepsilon(\bfr)$ and $\hat\bfmu(\bfr)$. The scattering features of such a medium is determined by the $3\times 3 $ matrix-valued functions,
	$\bfeta_\bfvarepsilon(\bfr):=\hat\bfvarepsilon(\bfr)-\bI$ and
	$\bfeta_\bfmu(\bfr):=\hat\bfmu(\bfr)-\bI$,
where $\bI$ is the $3\times 3$  identity matrix. The standard electromagnetic scattering theory \cite{TKD,newton} applies provided that, for $r:=|\bfr|\to\infty$, the entries of $\bfeta_\bfvarepsilon(\bfr)$ and $\bfeta_\bfmu(\bfr)$ tend to zero faster than $r^{-1}$. In particular, for $r\to\infty$, the electric field of the wave takes the form 
	$\bE(\bfr,t)=E_0 e^{-i\omega t}\left[e^{i\bk_{\rm i}\cdot\bfr}
	\hat\bE_{\rm i}+\frac{e^{ikr}}{r}\,\bF(\bk_{\rm s},\bk_{\rm i})\right]$,
where $E_0$ is a complex number, $\hat\bE_{\rm i}$ is the polarization vector for the incident wave, and $\bF(\bk_{\rm s},\bk_{\rm i})$ is a vector-valued function that plays the role of the scattering amplitude. In particular, given $\cR\subseteq\R^+$ and $\Omega\subseteq S^2$, the medium is invisible for incident wavenumbers $k$ ranging over $\cR$ and incidence directions $\hat\bk_{\rm i}$ belonging to $\Omega$ if and only if
	\begin{align}
	&\bF(\bk_{\rm s},\bk_{\rm i})=0~~{\rm for}~~k\in \cR~~{\rm and}~~\hat\bk_{\rm i}\in \Omega.
	\label{condi2-EM}
	\end{align}
This is equivalent to 	
	\be
	\sigma_T(\bk_{\rm i})=0~~{\rm for}~~k\in \cR~~{\rm and}~~\hat\bk_{\rm i}\in \Omega,
	\label{condi2-EM2}
	\ee
where $\sigma_T(\bk_{\rm i})$ denotes the total scattering cross-section \cite{TKD};
	$\sigma_T(\bk_{\rm i}):=\int_{S^{2}}d^{2}\hat\bk_{\rm s}\; \big|\bF(k\hat\bk_{\rm s},\bk_{\rm i})\big|^2$.

In Ref.~\cite{p176} we show that, for each $\alpha\in\R^+$, the first Born approximation provides the exact solution of the scattering problem for incident waves with wavenumber $k\leq\alpha$, if the following conditions hold.\\[6pt]
1. There are real numbers $a_\pm$ with $a_-<a_+$ such that 		
	$\bfeta_\bfvarepsilon(x,y,z)=\bfeta_\bfmu(x,y,z)=\bzero~~\for~~z\notin (a_-,a_+)$.\\[6pt]
2. The entries $\hat\varepsilon_{33}$ and $\hat\mu_{33}$ of $\hat\bfvarepsilon$ and $\hat\bfmu$ are bounded functions and their real parts have a positive lower bound, i.e., there are positive real numbers $m$ and $M$ such that $m\leq\RE[\hat\varepsilon_{33}(\bfr)]\leq |\hat\varepsilon_{33}(\bfr)|\leq M$ and $m\leq\RE[\hat\mu_{33}(\bfr)]\leq |\hat\mu_{33}(\bfr)|\leq M$, where ``$\RE$'' stands for the real part of its argument.\\[6pt]
3. There is a unit vector $\bu$ lying on the $x$-$y$ plane such that 
		\be
		\tilde{\bfeta}_\bfvarepsilon(\vec p,z)=		
		\tilde{\bfeta}_\bfmu(\vec p,z)=\bzero~~~~\for~~~~
		\bu\cdot\vec p\leq \alpha.
		\label{condi-EM}
		\ee
Conditions~1 and 2 are valid for a vast majority of realistic scattering setups. Note also that neither of these conditions nor Condition~3 requires the scattering medium to be isotropic, passive, or nondispersive.\footnote{In Supplementary Material we show that there is no theoretical inconsistency between these conditions and the temporal Kramers-Kronig relations imposed by frequency dispersion.}

We can express (\ref{condi-EM}) in the form
	\be
	\tilde{\bfeta}_\bfvarepsilon(\bp)=		
	\tilde{\bfeta}_\bfmu(\bp)=\bzero~~~~\for~~~~
	\bu\cdot\bp\leq \alpha.
	\label{condi-EM2}
	\ee
which is the electromagnetic analog of (\ref{condi-2Db}).\footnote{Notice however that here $\bu$ lies on the $x$-$y$ plane.} This suggests that our results on broadband directional invisibility for scalar waves extends to the scattering of electromagnetic waves by a medium satisfying Conditions 1 and 2. This is actually true, because the first Born approximation gives an expression for $\bF(\bk_{\rm s},\bk_{\rm i})$ that depends linearly on the entries of $\tilde{\bfeta}_\bfvarepsilon(\bk_s-\bk_i)$ and $\tilde{\bfeta}_\bfmu(\bk_s-\bk_i)$, \cite{newton}. In particular, $\bF(\bk_{\rm s},\bk_{\rm i})=\bzero$ whenever $\tilde{\bfeta}_\bfvarepsilon(\bk_s-\bk_i)=\tilde{\bfeta}_\bfmu(\bk_s-\bk_i)=\bzero$. This observation allows us to express the invisibility condition (\ref{condi2-EM}) in the form (\ref{condi-main}), where $\Pi$, $\Delta$, and $\cZ$ are respectively given by (\ref{Pi-def-3D}), (\ref{Delta-def-3D}), and 
$\cZ:=\left\{\bp\in\R^3\,\Big|\,\tilde{\bfeta}_\bfvarepsilon(\bp)=\tilde{\bfeta}_\bfmu(\bp)=\bzero\right\}$.
	
\begin{figure}
        	\begin{center}
        	\includegraphics[scale=.6]{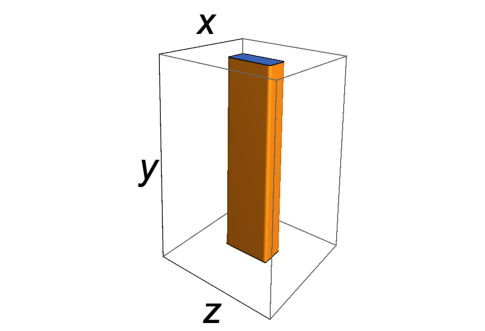}~\hspace{.1cm}\includegraphics[scale=.35]{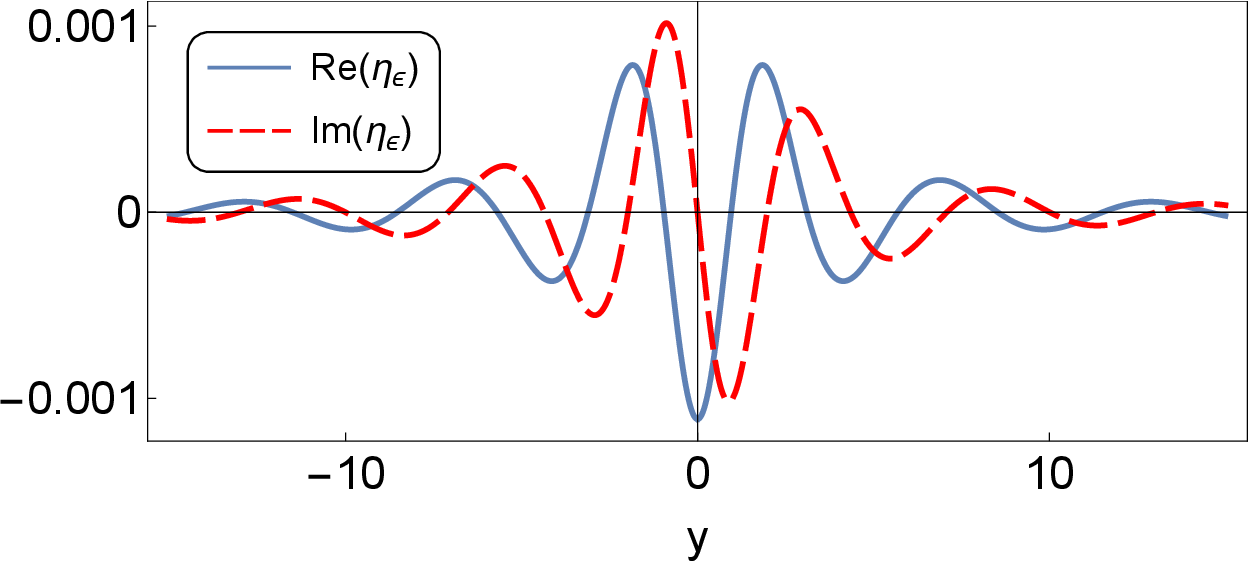}
        	\caption{Schematic view of the box confining the inhomogeneous part of the medium given by (\ref{eg-EM}), (\ref{eg-EM-2}), and (\ref{spec}) (on the left) and the plots of the real and imaginary parts of $\eta_\varepsilon:=\hat\varepsilon-1$ as a function of $y$ inside this box (on the right). Here we use units where $\alpha=1$.} 
        \label{fig2}
        \end{center}
        \end{figure}

As a specific example, consider a nonmagnetic isotropic medium with relative permittivity
	\be
	\hat\varepsilon(x,y,z):=1+\frac{\fz\,e^{i\alpha y}\chi_a(x)\chi_b(z)}{(y+i\ell)^{n+1}},
	\label{eg-EM}
	\ee
where $\fz$ is a real or complex coupling constant, and $a,b$, and $\ell$ are positive real parameters. Then 
	\begin{align}
	&\hat\bfmu=\bI, 
	&&\hat\bfvarepsilon=\hat\varepsilon\,\bI,
	&&\bfeta_\bfmu=\bzero,
	&&\bfeta_\bfvarepsilon=(\hat\varepsilon-1)\bI,
	\label{eg-EM-2}
	\end{align} 
and (\ref{condi-EM2}) holds for $\hat\bu=\hat\by$. The permittivity profile (\ref{eg-EM}) corresponds to an inhomogeneity confined to a box of infinite hight and rectangular base given by $|x|\leq \frac{a}{2}$ and $|z|\leq\frac{b}{2}$. See Fig.~\ref{fig2}.
This box consists of regions of gain and loss. Inside it, the amplitude of the inhomogeneity decays as $|y/\ell|^{-n-1}$ for $y\to\pm\infty$. Therefore, one can approximate $\hat\varepsilon(x,y,z)$ with
	\be
	\hat\varepsilon_L(x,y,z):=1+\frac{\fz\,e^{i\alpha y}\chi_a(x)\chi_{L}(y)\chi_b(z)}{(y+i\ell)^{n+1}},
	\label{eg-EM-approx}
	\ee
where $L$ is a positive real parameter much larger than $\ell$.  This corresponds to a scattering medium confined to the finite box given by $|x|\leq \frac{a}{2}$, $|y|\leq \frac{L}{2}$ and $|z|\leq\frac{b}{2}$. For 
	\begin{align}
	&n=1, &&\fz=0.01\alpha^{-2}, && a=\alpha^{-1}, && b=2\alpha^{-1}, && \ell=3\alpha^{-1},
	\label{spec}
	\end{align} 
and $L=15\alpha^{-1}$,  we  have $|\hat\varepsilon(x,y,z)-\hat\varepsilon_L(x,y,z)|<4.3\times 10^{-5}$.

To demonstrate the broadband directional invisibility of the scattering medium $\sM$ defined by (\ref{eg-EM}), (\ref{eg-EM-2}), and (\ref{spec}), we examine the behavior of its total cross-section $\sigma_T(\bk_{\rm i})$ for cases where $\bk_{\rm i}\cdot\hat\bx=0$, i.e., $\bk_{\rm i}=(0,k_y,k_z)$ for some $k_y,k_z\in\R$. In other words we explore broadband directional invisibility corresponding to $\Omega\subseteq S^1_{0}$ where
$S^1_{0}:=\{(0,\sin\theta_0,\cos\theta_0)\;|\;\theta_0\in[0,360^\circ)\}$. 
For $k\leq\alpha$, where the first Born approximation is exact, $\sigma_T(\bk_{\rm i})$ is proportional to 
	$\breve\sigma_T(\bk_{\rm i}):=\int_{S^{d-1}}d^{d-1}\hat k_{\rm s}\; \big|\tilde\eta_{\varepsilon}(k\hat\bk_{\rm s}-\bk_{\rm i})\big|^2$,
where $\tilde\eta_{\varepsilon}(\bp)$ stands for the 3D Fourier transform of $\eta_{\varepsilon}(\bfr):=\hat\varepsilon(\bfr)-1$.
Therefore the medium displays invisibility for incident wavenumbers $k\leq\alpha$ if and only if $\breve\sigma_T(\bk_{\rm i})=0$. 

\begin{figure}
        \begin{center}
        \includegraphics[scale=.54]{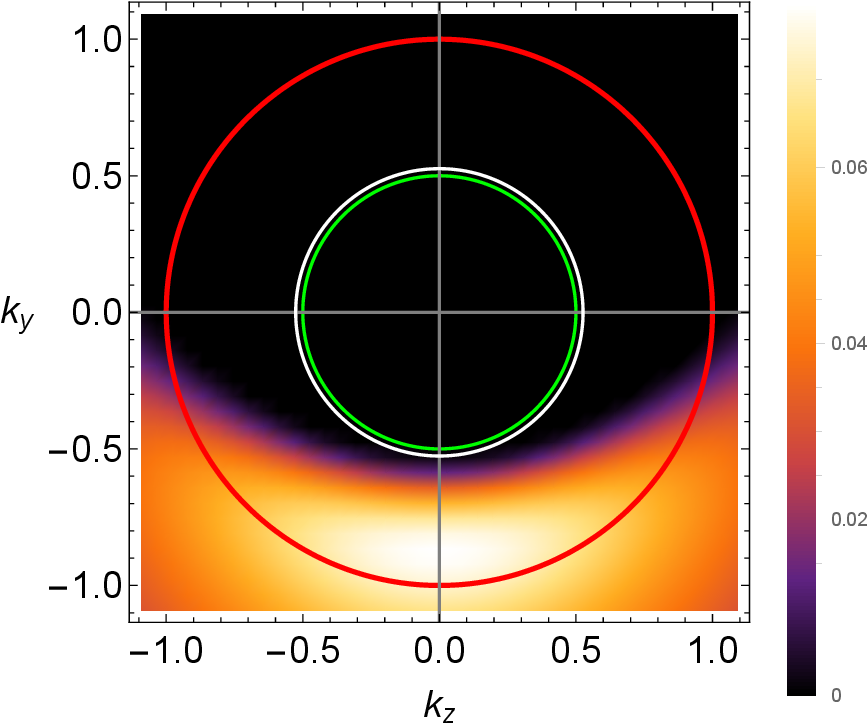}~~~~~
        \includegraphics[scale=.53]{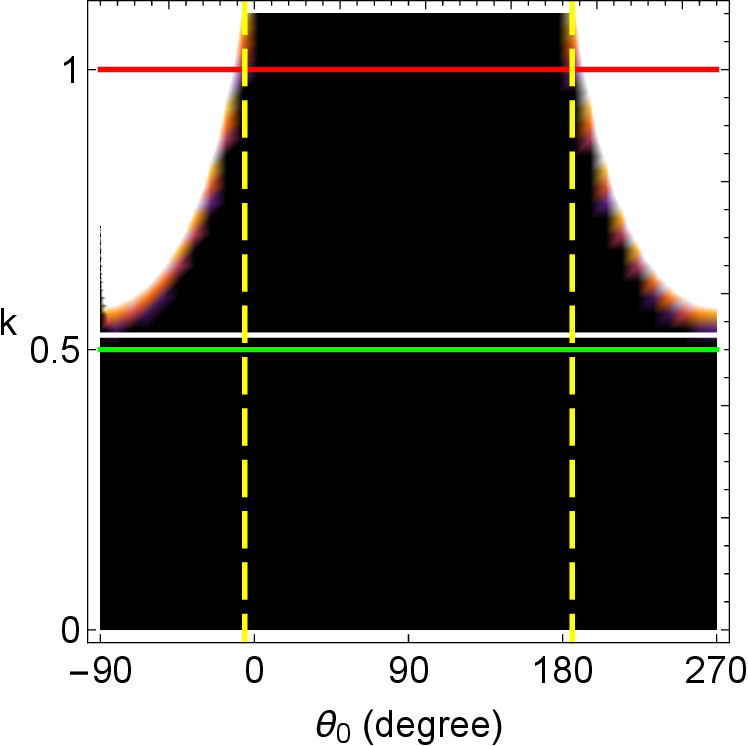}
        \caption{Density plots of $\breve\sigma_T(0,k_y,k_z)$ (on the left) and $\breve\sigma_T(0,k\sin\theta_0,k\cos\theta_0)$ (on the right) for the medium given by (\ref{eg-EM}), (\ref{eg-EM-2}), and (\ref{spec}) in units where $\alpha=1$. The green, white, and red circles are respectively the boundaries of the discs $\sD_{\frac{\alpha}{2}}$, $\sD_{\rho_\star }$, and $\sD_\alpha$. The ranges $(-90^\circ,90^\circ)$ and $(90^\circ,270^\circ)$ for $\theta_0$ correspond to situations where the source of the incident wave lies on the planes $z=-\infty$ and $z=+\infty$, respectively. The medium displays directional invisibility in the black region lying between the white and red circles  (respectively lines) in the plot on the left (respectively right). The dashed yellow lines correspond to $\theta_0=-\beta$ and $\theta_0=180^\circ+\beta$. } 
        \label{fig3}
        \end{center}
        \end{figure}

Figure~\ref{fig3} shows density plots of $\breve\sigma_T(0,k_y,k_z)$ and $\breve\sigma_T(0,k\sin\theta_0,k\cos\theta_0)$ for the medium $\sM$ given by (\ref{eg-EM}), (\ref{eg-EM-2}), and (\ref{spec}). Let $\sD_\rho$ denote the disc of radius $\rho$ defined by $\sqrt{k_z^2+k_y^2}\leq\rho$ in the $k_{z}$-$k_{y}$ plane. In the plot of $\breve\sigma_T(0,k_y,k_z)$ (respectively $\breve\sigma_T(0,k\sin\theta_0,k\cos\theta_0)$), $\sD_\alpha$ (respectively the half-plane $k\leq\alpha$) signifies the region where the first Born approximation is exact and our results apply. The region where $\breve\sigma_T$ vanishes is colored in black. $\sM$ is invisible for wave vectors belonging to the black region lying in $\sD_\alpha$ (respectively below the line $k=\alpha$). We use ${Q}$ to refer to this invisibility region. The largest disc centered at the origin that is contained in ${Q}$ is the region in which the system displays invisibility for $\Omega=S^1_0$. The radius of this disc which we denote by $\rho_\star $ turns out to be $0.5255\,\alpha$.\footnote{The fact that $\rho_\star >\frac{\alpha}{2}$ is consistent with the omnidirectionally invisibility of the system for $k\leq\frac{\alpha}{2}$.} $\sD_\alpha$ includes a region (colored in different shades of orange) where the system scatters the incident wave. This shows that for every subinterval $\cR:=[\rho_-,\rho_+]$ of $(\rho_\star ,\alpha]$, there is a set $\Omega_{\rho_-,\rho_+}$ of directions $\hat\bk_{\rm i}$ of the incident wave vector such that the system is invisible for $k\in\cR$ and $\hat\bk_{\rm i}\in\Omega_{\rho_-,\rho_+}$, but $\Omega_{\rho_-,\rho_+}\neq S_0^1$. Because $\Omega_{\rho_-,\rho_+}\subseteq S_0^1\subseteq S^2$, this implies that $\Omega_{\rho_-,\rho_+}\neq S^2$, i.e., the medium displays broadband directional invisibility. For example, choosing the optimal values of $\rho_\pm$, i.e., $\rho_-=\rho_\star=0.5255\,\alpha $ and $\rho_+=\alpha$, we find that it is invisible for $k\in [\rho_\star,\alpha]$ and $\hat\bk_{\rm i}$ belonging to 
$\Omega_{\rho_\star,\alpha}=\{(0,\sin\theta_0,\cos\theta_0)|-\beta\leq\theta_0\leq 180^\circ+\beta\}$, 
where $\beta:=5.637^\circ$. 
        

In conclusion, we would like to recall that during the past ten years or so, there has been a growing interest in the utility of material with balanced gain and loss regions in achieving nonreciprocal invisibility. Great majority of the results obtained in this direction are however restricted to (effectively) one-dimensional systems. In the present article we have given a precise and systematic description of broadband directional invisibility for scalar and electromagnetic waves. We have then employed our recent results on the exactness of the first Born approximations to provide explicit conditions on a scattering potential or linear scattering medium that ensure their broadband directional invisibility. 

The concrete examples of potentials and permittivity profiles we have constructed for the purpose of achieving broadband directional invisibility correspond to scatterers with regions of gain and loss. In this sense our findings may be considered as an extension of the results on broadband unidirectional invisibility in one dimension \cite{horsley-2015,longhi-2015,horsley-longhi}. {A major difference is that our approach allows us to achieve invisibility in a finite wavenumber spectrum $\sR$. It also offers vast freedom in the choice of directions for which we can maintain invisibility of the scatterer.  Another notable feature of this approach is that it applies for wavenumbers below a prescribed value $\alpha$, which makes it particularly effective at low frequencies.} This is in sharp contrast with the approach of Refs.~\cite{hayran-2018a,hayran-2018b} which is reliable only at sufficiently high frequencies where the first Born approximation is valid (but not exact).  

An important aspect of the present work which requires particular attention is its practical implementation {where frequency dispersion is expected to restrict the choice of $\alpha$ and size of $\sR$.} The experimental realizations \cite{jiang-2017,ye-2017} of the full-band invisibility reported in Refs.~\cite{horsley-2015,longhi-2015} provide encouragement towards experimental verifications of our results on broadband directional invisibility.\vspace{6pt}

\noindent Supplementary Material: Effects of frequency dispersion on the implementation of our approach to broadband directional invisibility for electromagnetic waves. \vspace{6pt}

\noindent Acknowledgements:
This work has been supported by the Scientific and Technological Research Council of T\"urkiye (T\"UB\.{I}TAK) in the framework of the project 120F061 and by Turkish Academy of Sciences (T\"UBA).

\np

\begin{center}
\noindent{\large\bf Supplementary Material: Implications of frequency dispersion}\vspace{12pt}
\end{center}


The conditions we have found for the theoretical realization of broadband directional invisibility require the linear media to involve regions of gain and loss. Because the gain/loss profile of the medium is sensitive to the frequency of the incident wave, we wish to examine the compatibility of our results with the presence of frequency dispersion. For this reason, we consider situations where the permittivity and permeability tensors of the medium depend  on the angular frequency of the wave $\omega:=ck$, and that they respect the temporal Kramers-Kronig relations. 

To simplify our analysis, we focus our attention on non-magnetic isotropic media. We can then state the temporal Kramers-Kronig relations  in the form
	\begin{align}
	&\eta_\varepsilon(x,y,z,\omega)=\frac{1}{i\pi} {\rm P}\!\!
	\int_{-\infty}^\infty d\omega'\:\frac{\eta_\varepsilon(x,y,z,\omega')}{\omega'-\omega},
	 \label{KK}
	\end{align}
where $\eta_\varepsilon(x,y,z,\omega)=\hat\varepsilon(x,y,z,\omega)-1$ is 
the electric susceptibility, and ${\rm P}$ stands for the principle value of the integral. Equation~(\ref{KK}) identifies $\eta_\varepsilon(x,y,z,\omega)$ with a function of $\omega$ that is analytic in the upper complex $\omega$ half-plane ($\IM(\omega)>0$). In general it has poles in the lower complex $\omega$ half-plane ($\IM(\omega)<0$). It is clear that this condition does not violate conditions 1 and 2 of the exactness of the first Born approximation that we employ in our analysis, namely
	\begin{itemize}
	\item[1.] There are real numbers $a_\pm$ with $a_-<a_+$ such that for all $\omega\in\R^+$,
	\be
	\eta_\varepsilon(x,y,z,\omega)=0~~\for~~z\notin (a_-,a_+).
	\nn
	\ee
	\item[2.] There are positive real numbers $m$ and $M$ such that for all $\omega\in\R^+$,
	\begin{align}
	&m\leq\RE[\hat\varepsilon(x,y,z,\omega)]\leq |\hat\varepsilon(x,y,z,\omega)|\leq M.
	\nn
	\end{align}
	\end{itemize}
Next, we consider Condition~3, namely
	\begin{itemize}
	\item[3.] There is a unit vector $\bu$ lying on the $x$-$y$ plane such that 
for all $\omega\in\R^+$,
		\be
		\tilde{\eta}_\varepsilon(\vec p,z,\omega)=0~~~~\for~~~~
		\bu\cdot\vec p\leq \alpha.
		\label{condi-EM}
		\ee
	\end{itemize}
As we explain in our paper, we can choose our coordinate system such that $\bu=\hat\by$. In this case (\ref{condi-EM}) takes the form
	\be
	\tilde{\eta}_\varepsilon(p_x,p_y,z,\omega)=0~~~~\for~~~~
	p_y\leq \alpha.
	\label{condi-EM-2}
	\ee
		
Let $\breve\eta_\varepsilon(x,y,z,\tau)$ denote the Fourier transform of $\eta_\varepsilon(x,y,z,\omega)$ with respect to $\omega$, i.e.,
	\[\breve\eta_\varepsilon(x,y,z,\tau):=\int_{-\infty}^\infty d\omega\: e^{-i\omega\tau}\eta_\varepsilon(x,y,z,\omega).\]
Then (\ref{KK}) is equivalent to
	\be
	\breve\eta_\varepsilon(x,y,z,\tau)=0~~~~\for~~~~\tau<0,
	\label{KK2}
	\ee
and consequently
	\be
	\tilde{\breve\eta}_\varepsilon(p_x,p_y,z,\tau)=0~~~~\for~~~~\tau<0,
	\label{KK3}
	\ee
where a tilde stands for the Fourier transform with respect to $x$ and $y$. Equation~(\ref{KK3}) is the form of the temporal Kramers-Kronig relations that we wish to compare with Condition 3 of our paper, namely (\ref{condi-EM-2}). Evaluating the Fourier transform of both sides of the equation in (\ref{condi-EM-2}) with respect to $\omega$, we can write it in the form
	\be
	\breve{\tilde{\eta}}_\varepsilon(p_x,p_y,z,\tau)=0~~~~\for~~~~
	p_y\leq \alpha.
	\nn
	\ee
Because $\tilde{\breve\eta}_\varepsilon=\breve{\tilde\eta}_\varepsilon$, we can identify this condition with 
	\be
	\tilde{\breve\eta}_\varepsilon(p_x,p_y,z,\tau)=0~~~~\for~~~~
	p_y\leq \alpha.
	\label{condi-EM-3}
	\ee
	
\begin{figure}
        \begin{center}
        \includegraphics[scale=.55]{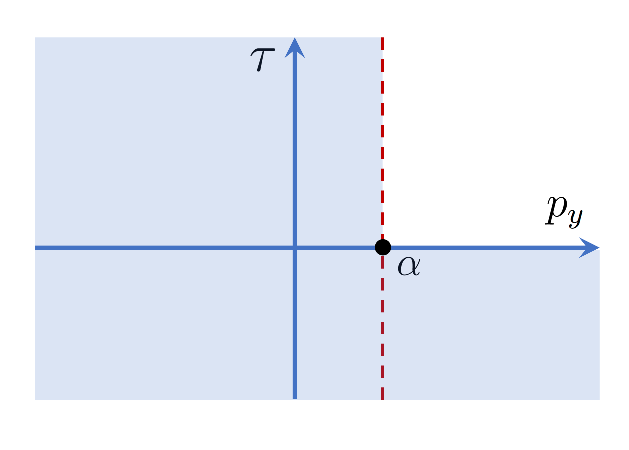} 
        \caption{Union $\Delta$ of the half-planes given by $p_y<\alpha$ and $\tau<0$ in the  $p_y$-$\tau$ (colored in light blue). The red dashed line corresponds to $p_y=\alpha$.}
        \label{picMS1}
        \end{center}
        \end{figure}

Our results on the exactness of the first Born approximation and broadband directional invisibility would be in conflict with the temporal Kramers-Kronig relations if and only if Condition (\ref{condi-EM-3}) violates (\ref{KK3}). These conditions hold for the functions $\tilde{\breve\eta}_\varepsilon$ that vanish in the set 
	\[S:=\{(p_z,p_y,z,\tau)\in\R^4~|~(p_y,\tau)\in\Delta\},\] 
where $\Delta$ is the union of the half-planes given by $p_y<\alpha$ and $\tau<0$ in the $p_y$-$\tau$ plane (See Fig.~\ref{picMS1}.) Because we can easily find explicit examples of functions vanishing in $S$, Condition (\ref{condi-EM-3}) does not violate (\ref{KK3}). {This shows that the presence of frequency dispersion does not obstruct conditions we found for broadband directional invisibility. However, in specific experimental implementations of our approach, it will not be possible to satisfy both (\ref{KK3}) and (\ref{condi-EM-3}) for an arbitrary choice of $\alpha$. This shows that frequency dispersion can impose severe restrictions on the allowed values of $\alpha$ and the size of the wavenumber domain $\sR$ for directional invisibility.} The results of Ref.~8 on the experimental realizations of a permittivity profile satisfying one-dimensional analogs of (\ref{KK3}) and (\ref{condi-EM-3}) with $\alpha=0$ suggest that fulfilling these restrictions should not be impossible. 

\ed